\documentclass[aps,preprint,showpacs]{revtex4}

\begin{document}

\title{Multicolor vortex solitons in two-dimensional photonic lattices}
\author{Zhiyong Xu$^{1}$, Yaroslav V. Kartashov$^{1,2}$, Lucian-Cornel Crasovan$^{1,3}$,
Dumitru Mihalache$^{3}$, and Lluis Torner$^{1}$}

\affiliation{$^{1}$ICFO-Institut de Ciencies Fotoniques, and
Department of Signal Theory and Communications, Universitat
Politecnica de Catalunya, 08034 Barcelona,
Spain}

\affiliation{$^{2}$Physics Department, M. V. Lomonosov Moscow
State University, 119899 Moscow, Russia}

\affiliation{$^{3}$Institute of Atomic Physics, Department of
Theoretical Physics, P.O. Box MG-6, Bucharest, Romania}

\begin{abstract}
We report on the existence and stability of multicolor lattice
vortex solitons constituted by coupled fundamental frequency and
second-harmonic waves in optical lattices in quadratic nonlinear
media. It is shown that the solitons are stable almost in the
entire domain of their existence, and that the instability domain
decreases with the increase of the lattice depth. We also show the
generation of the solitons, and the feasibility of the concept of
{\it lattice soliton algebra}.
\end{abstract}


\maketitle

Localized structures, especially solitons, play a crucial role in
many branches of the nonlinear science. Over the past several
decades, the existence and unique properties of spatial, temporal
and spatiotemporal optical solitons in homogeneous cubic and
quadratic nonlinear media have been studied both theoretically and
experimentally (for detailed reviews,
see\cite{Kivshar0,Stegeman0,Buryak0}). Solitons arise as a result
of the balance between linear diffracting and/or dispersing
properties of the medium and a nonlinear mechanism responsible for
focusing /defocusing. One important subject of study is the
generation of nonlinear modes with a nontrivial phase, such as
vortices. In optics, vortices are associated with screw phase
dislocations nested in light beams \cite{Soskin}. Here we are
interested in vortices with a bright shape, i.e., dislocations
nested in finite-size beams \cite{ring-vortices}. In homogeneous
pure cubic and quadratic uniform media such ring-shaped vortex
solitons suffer azimuthal instabilities which have been observed
experimentally in different settings \cite{AMI-observations}. They
can be made stable in media with competing nonlinearities
\cite{competing}, and in media with refractive index modulations
that we address here.

Propagation of optical radiation in media with transverse
refractive index modulation differs considerably from the
propagation in uniform media. Localized structures in such
periodic media, termed discrete or lattice solitons, do exist and
exhibit a rich variety of topologies. Since the theoretical
prediction of discrete optical solitons in 1988
\cite{Christodoulides1} they keep attracting a growing interest,
in part because of their potential for all-optical switching and
routing phenomena \cite{Christodoulides2,ole1,Yar1}. The
intermediate regime between continuous and discrete solitons
\cite{Scharf,Cohen} constituted by continuous nonlinear media with
an imprinted transverse modulation of the refractive index, has
been shown recently to offer a variety of new opportunities. The
concept behind this regime might be termed {\it tunable
discreteness}, the strength of modulation being the parameter that
tunes the system properties from continuous to discrete. In this
context, wave dynamics is governed by the interplay between
optical tunnelling to adjacent sites and nonlinearity. This kind
of lattice solitons have been observed recently in two-dimensional
(2D) photorefractive optical lattices
\cite{Efremidis,Fleischer1,Fleischer2,Neshev0,Yar3}.

Besides the study of the existence and stability properties of
fundamental (ground state) modes in the nonlinear media with
periodic potentials an intriguing question is whether the action
of the confining potential can permit the formation of vorticity
carrying localized structures. Theoretical works showed that such
complex localized structures, i.e., {\it lattice vortex solitons}
exist when an optical lattice acts on a Kerr or photorefractive
nonlinear crystal \cite{Malomed,Yang1,Musslimani}. Recently,
theoretical expectations were indeed confirmed experimentally by
two independent groups \cite{Neshev,Fleischer3}. During the last
years various families of solitons in arrays of weakly coupled
waveguides made with quadratic nonlinear media have been also
investigated \cite{Sukhorukov,Bang2,Pesche,Kobyakov,Malomed2,Xu},
and observed for the first time recently \cite{george-PRL}.
One-dimensional (1D) multicolor solitons in lattices with tunable
strength have been also studied recently, and their potential
applications for packing and steering single solitons have been
investigated \cite{Yaroslav1}. Two-dimensional geometries might
support even robust soliton ensembles with phase dislocations, a
problem not analyzed so far, and which is the aim of this work.

We thus report on the existence and stability of such multicolor
lattice vortex solitons, which comprise four main humps arranged
in a square configuration. It is shown that the lattice vortex
solitons are stable almost in the entire domain of their existence
except for a narrow region near the cutoff, and that the
instability domain decreases with the increase of the lattice
depth. We also investigate the possibility of their dynamical
generation \cite{Silvia} from Gaussian-type input beams with
nested vorticities.

We study the system of coupled nonlinear equations that describe
the interaction between the fundamental frequency (FF) and
second-harmonic (SH) waves under conditions for type I
second-harmonic generation in bulk materials in the absence of
Poynting vector walk-off:
\begin{eqnarray}
&&i\frac{\partial q_1}{\partial
\xi}=\frac{d_{1}}{2}\left(\frac{\partial^{2}q_{1}}{\partial\eta^{2}}+\frac{\partial^{2}q_{1}}{\partial\zeta^{2}}\right)
-q_1^{*}q_2\exp \left( -i\beta\xi\right)-pR\left(\eta,\zeta\right)q_{1}   \nonumber \\
&&i\frac{\partial q_2}{\partial
\xi}=\frac{d_{2}}{2}\left(\frac{\partial^{2}q_{2}}{\partial\eta^{2}}+\frac{\partial^{2}q_{2}}{\partial\zeta^{2}}\right)
-q_1^{2}\exp \left(
i\beta\xi\right)-2pR\left(\eta,\zeta\right)q_{2} ,
\end{eqnarray}
where
$q_1=\left(2k_{1}/k_{2}\right)^{1/2}[2\pi\omega_{0}^{2}\chi^{(2)}r_{0}^{2}/c^{2}]A_{1}$
and $q_2=[2\pi\omega_{0}^{2}\chi^{(2)}r_{0}^{2}/c^{2}]A_{2}$
represent the normalized complex amplitudes of the FF and SH
fields, $k_{1}=k(\omega_{0})$,
$k_{2}=k(2\omega_{0})\approx2k_{1}$, $r_{0}$ is the transverse
scale of the input beams, $\eta=x/r_{0}$, $\zeta=y/r_{0}$,
$\xi=z/(k_{1}r_{0}^{2})$, $\beta=(2k_{1}-k_{2})k_{1}r_{0}^{2} $ is
the phase mismatch, $d_{1}=-1$, $d_{2}=-k_{1}/k_{2}\approx-1/2$,
and $p=2\pi\omega_{0}^{2}\delta\chi^{(1)}r_{0}^{2}/c^{2}$ is the
lattice depth.
 The function $R(\eta,\zeta)=\cos(2\pi\eta/T)\cos(2\pi\zeta/T)$
describes the transverse refractive index profile, where $T$ is
the modulation period. The system (1) admits several conserved
quantities, including the energy flow
\begin{eqnarray}
&&U=\int\int_{-\infty}^{+\infty}(|q_{1}|^{2}+|q_{2}|^{2})d\eta
d\zeta,
\end{eqnarray}and the Hamiltonian
\begin{eqnarray}
&&H=\int\int_{-\infty}^{+\infty}[-\frac{d_{1}}{2}|\nabla
q_{1}|^{2}-\frac{d_{2}}{4}|\nabla q_{2}|^{2}
\nonumber \\
&&-\frac{1}{2}(q_1^{*})^{2}q_2\exp \left(
-i\beta\xi\right)-\frac{1}{2}q_1^{2}q_2^{*}\exp \left(
i\beta\xi\right)
\nonumber \\
&&+\frac{\beta}{2}|q_{2}|^{2}-pR(\eta,\zeta)|q_{1}|^{2}-pR(\eta,\zeta)|q_{2}|^{2}
]d\eta d\zeta,
\end{eqnarray}
where $\nabla =\mathbf{{e_{\eta}}}(\partial /\partial
\eta)+\mathbf{{e_{\zeta}}}(\partial /\partial \zeta)$, and
$\mathbf{{e_{\eta}}}$, $\mathbf{{e_{\zeta}}}$ are unity vectors
along $\eta$ and $\zeta$ axes.

We searched for the stationary solutions in the form $q_{1}=(u_{1}
(\eta,\zeta)+iv_{1}(\eta,\zeta))\exp(ib_{1}\xi)$ and $q_{2}=(u_{2}
(\eta,\zeta)+iv_{2}(\eta,\zeta))\exp(ib_{2}\xi)$, where
$u_{1,2}(\eta,\zeta)$ and $v_{1,2}(\eta,\zeta)$ are real
functions, and $b_{1,2}$ are real propagation constants that
verify $b_{2}=\beta+2b_{1}$. Substitution of the above expressions
into Eq. (1) yields the following system of equations for the
soliton profiles $u_{1,2}$ and $v_{1,2}$
\begin{eqnarray}
&&\frac{d_{1}}{2}\left(\frac{\partial^{2}u_{1}}{\partial
\eta^{2}}+\frac{\partial^{2}u_{1}}{\partial \zeta^{2}}\right)-
u_{1}u_{2}-v_{1}v_{2}+b_{1}u_{1}-pR(\eta,\zeta)u_{1} = 0  \nonumber \\
&&\frac{d_{1}}{2}\left(\frac{\partial^{2}v_{1}}{\partial
\eta^{2}}+\frac{\partial^{2}v_{1}}{\partial \zeta^{2}}\right)-
u_{1}v_{2}+v_{1}u_{2}+b_{1}v_{1}-pR(\eta,\zeta)v_{1} = 0  \nonumber \\
&&\frac{d_{2}}{2}\left(\frac{\partial^{2}u_{2}}{\partial
\eta^{2}}+\frac{\partial^{2}u_{2}}{\partial \zeta^{2}}\right)-
u_{1}^{2}+v_{1}^{2}+b_{2}u_{2}-2pR(\eta,\zeta)u_{2} = 0  \nonumber \\
&&\frac{d_{2}}{2}\left(\frac{\partial^{2}v_{2}}{\partial
\eta^{2}}+\frac{\partial^{2}v_{2}}{\partial \zeta^{2}}\right)-
2u_{1}v_{1}+b_{2}v_{2}-2pR(\eta,\zeta)v_{2} = 0.
\end{eqnarray}

We solved the system of coupled equations (4) numerically by using
a standard relaxation method. The lattice vortex soliton families
are one-parameter families defined by the propagation constant
$b_{1}$ for any given period of the modulation $T$, lattice depth
$p$ and phase mismatch $\beta$. Since one can use scaling
transformations
$q_{1,2}(\eta,\zeta,\xi,\beta,p)\rightarrow\chi^{2}q_{1,2}
(\chi\eta,\chi\zeta,\chi^{2}\xi,\chi^{2}
\beta,\chi^{2}p)$ to obtain various families of solitons from a
given family, we have selected the transverse scale $r_{0}$ such
that the modulation period is given by $T=\pi/2$ and then we have
varied $b_{1}$, $\beta$, and $p$.

The simplest vortex soliton with unit topological charge in
two-dimensional periodic lattice is shown in Fig. 1. It comprises
four main humps arranged in a square configuration with a
stair-like phase structure that is topologically equivalent to the
phase of a conventional vortex in uniform medium (see Fig.
1(b,d)). The positions of the soliton intensity maxima almost
coincide with the positions of the local maxima of the lattice.
Note that the singularity of these vortex solitons is centered
between four lattice sites, that is they belong to the class of
the, so called, {\it off-site} vortex soliton. In the model we
investigated here there exist also a family of {\it on-site}
vortex solitons (not shown here). In that case the phase
singularity is centered on a lattice site
\cite{Yang1,Neshev,Fleischer3}. We will restrict ourselves here to
the case of the off-site vortex solitons. It is interesting to
note that these stationary structures somehow resembles the
4-soliton molecules carrying orbital angular momentum that were
investigated in a variety of nonlinear media in both
two-dimensional and three-dimensional geometries
\cite{sol_molec1,sol_molec2,sol_molec3}. The typical stair-like
phase distribution in the case of the above mentioned soliton
molecules is clearly seen in panels (b) and (d) of Fig. 1 for the
lattice vortex solitons. We want to mention that the 1D quadratic
waveguides were also shown to support various families of
multipeaked solitons, which display combinations of in-phase and
out-of-phase odd solitons, the latter ones with $\pi$ phase jumps
between neighbor solitons \cite{Yaroslav1}.

It should be noted that at low powers (small $b_{1}$) the lattice
vortex solitons are quite wide and spread out over many lattice
sites (Fig. 1(a)), while at high powers the energy is mainly
localized within the corresponding four peaks (Fig. 1(c)). We did
not find four-hump structures with higher topological charges (2
or more), and all other higher-order stationary structures we have
found (for example, lattice vortex solitons with eight humps) were
found to be unstable on propagation. Thus, in this work we will
restrict ourselves to the study of the properties of simplest
four-hump lattice vortex solitons.

In order to characterize the families of lattice vortex solitons
we have calculated the energy flows associated with these
stationary solutions as well as their existence domains for given
lattice depth $p$ and phase mismatch $\beta$. As a general rule,
the energy flow of the four-hump lattice vortex solitons is a
non-monotonic function of the propagation constant (even if this
cannot be seen directly from Fig. 2(a) without zooming). Note that
in Fig. 2(a,d) on the abscise we have plotted the difference
$(b_{1}-b_{co})$ between the propagation constant $b_{1}$ and
cutoff value $b_{co}$. The cutoff value depends on both the phase
mismatch $\beta$ and lattice depth $p$. For example at $p=8$
cutoff is given by $b_{co}=6.065$ for $\beta=-6$, while
$b_{co}=1.23$ for $\beta=6$. The cutoff $b_{co}$ is a
non-monotonic function of the lattice depth $p$ [Fig. 2(b)]. It
tends to infinity at $p\rightarrow 0$ and $p\rightarrow +\infty$.
One can see from Fig. 2(c) that in the presence of the lattice the
dependence $b_{co}(\beta)$ differs from that for quadratic
solitons in continuous media: $b_{co}(\beta)={\rm
max}\{-\beta/2,0\}$. Thus at $\beta\rightarrow -\infty$ cut-off is
approximately given by $(\beta-\beta_{0})/2$, while at
$\beta\rightarrow +\infty$ one has $b_{co}=b_{0}$, where
$\beta_{0}$ is the mismatch shift due to the lattice. Both $b_{0}$
and $\beta_{0}$ growth with $p$. This property holds also for the
one-dimensional lattice solitons in quadratic nonlinear media
\cite{Yaroslav1}.

The periodic refractive index modulation affects also the energy
sharing between FF and SH waves. For example, at a given phase
mismatch $\beta$, the fraction of the total energy flow carried by
the SH wave increases with increase of the lattice depth.
Moreover, near the cut-off, the SH wave spreads over more lattice
sites than the FF beam. As in the case of a uniform media, in the
lattice with fixed depth $p$ the part of energy flow carried by
the SH wave decreases with increase of phase mismatch $\beta$.

To investigate the stability of the lattice vortex solitons, we
have performed extensive numerical simulations of the evolution
dictated by  Eq. (1) with the input conditions
$q_{1}(\xi=0)=(u_{1}(\eta,\zeta)+iv_{1}(\eta,\zeta))(1+\rho_{1}(\eta,\zeta))$
and
$q_{2}(\xi=0)=(u_{2}(\eta,\zeta)+iv_{2}(\eta,\zeta))(1+\rho_{2}(\eta,\zeta))$,
where $u_{1,2}$ and $v_{1,2}$ are the exact solutions of Eq. (4)
and $\rho_{1,2}$ are random functions with Gaussian distribution
and variance $\sigma_{noise}^{2}=0.01$. We have propagated the
perturbed four-hump lattice vortex solitons over thousands of
units for various values of the physical parameters involved
($\beta$, $p$, $U$).

Our simulations show that there exists a narrow instability band
near the propagation constant cutoff $b_{co}$ for vortex solitons,
but above certain critical value of propagation constant they
appear to become free of instability. We have found that the width
of instability domain of lattice vortex solitons decreases with
increase of the depth of the lattice $p$ [Fig. 2(d)]. For example,
as depicted in Fig. 2(d), the width of instability domain on
propagation constant for $p=12$ is approximately given by $0.24$,
while for $p=8$ it is $0.35$.

A few representative decay scenarios for the unstable four-hump
lattice vortex solitons with unit topological charge  are shown in
Fig. 3. In the row (a) of Fig. 3 we show the typical decay of the
unstable vortex soliton in the vicinity of the cutoff on
propagation constant $b_{co}$. The initial energy of the localized
structure is spreading out during evolution across the whole
lattice and the vortex soliton disappears. Notice that this type
of instability develops exponentially. In the rest part of the
instability domain located closer to critical value of the
propagation constant, we encountered oscillatory-type instability.
Upon development of this instability vortex soliton transforms
into a fundamental (ground state) lattice soliton that is the most
robust and energetically stable state of the system (see row (b)
of Fig. 3), through increasing oscillations of four intensity
maxima of the vortex.

One of the important results of this study is that the lattice
vortex soliton becomes completely stable when its propagation
constant exceeds a critical value $b_{cr}$, i.e., almost in the
entire existence domain (see Fig. 2(d)). In row (c) of Fig. 3 we
have plotted, for the sake of illustration,
 the initial and the final (after 500 propagation units) intensity
distributions of a stable lattice vortex soliton. Comparing to the
soliton molecules investigated in bulk nonlinear media, which were
shown to be {\it metastable} physical objects under suitable
conditions, we conclude that, as expected on physical grounds, the
effect of the two-dimensional lattice is to arrest the rotation of
the soliton molecule and thus to assure the complete stabilization
of the soliton complex. Since lattice causes strong azimuthal
modulation of the vortex soliton, lattice removal results in
complete soliton decay into four filaments, as shown in Fig. 4.
Escape angles of filaments decrease with increase of input energy
flow of vortex soliton.

To understand lattice vortex solitons generation from a radially
symmetric input beam carrying a screw phase dislocation nested in
the center and to show that different sets of output solitons can
be obtained with different combinations of topological charges and
shapes of the input beams we performed a comprehensive set of
simulations of Eq. (1) with the input conditions corresponding to
Gaussian beams with a phase dislocation nested in the center:
\begin{eqnarray}
&&q_{1}(\xi=0,r,\varphi)=Ar^{|m_{1}|} \exp(im_{1}\varphi) \exp(-r^{2}/w_{1}^{2}) \nonumber \\
&&q_{2}(\xi=0,r,\varphi)=Br^{|m_{2}|} \exp(im_{2}\varphi)
\exp(-r^{2}/w_{2}^{2}) ,
\end{eqnarray}
where $r=(\eta^{2}+\zeta^{2})^{1/2}$ is the radius, $\varphi$ is
the azimuthal angle, $A$ and $B$ are amplitudes of FF and SH waves
, $w_{1}$ and $w_{2}$ are beam widths. Below we set the width
$w_{1}=w_{2}=1$ and suppose that topological charge of FF wave is
given by $m_{1}=1$.

First, we consider non-seeded vortex soliton generation at $B=0$.
At low input powers both input FF wave and generated SH wave
exhibit complete diffraction, and input beam energy is
redistributed between many lattice sites. With the increase of
input energy flow (i.e. by increasing $A$ in Eqs. (5)) the
generation of lattice vortex soliton with unit topological charge
becomes possible as shown in Fig. 5. Lattice soliton generation is
accompanied with energy radiation (Fig. 5(c) and (d)) but the
ratio between radiative losses and the output soliton energy flow
decreases with increase of input energy flow.

In the case of seeded SH generation ($B \neq 0$ and $B \ll A$),
the output field distribution can be controlled by the input
topological charge of SH wave. For $m_{1}=1$ vortex soliton
generation is possible only for the vorticity-matched case when
$m_{2}=2$, while all other values of $m_{2}$ correspond to
formation of trivial-phase soliton distributions, whose structure
is dictated by lattice symmetry and energy exchange between FF and
SH waves at the initial stage of propagation. Some representative
output distributions are shown in Fig. 6. These plots show that
the concept of "soliton algebra" previously explored in
homogeneous media \cite{IBU}, does also apply in the presence of
lattices, offering interesting opportunities for controlling the
soliton dynamics.

In summary, we have shown that periodic lattices imprinted in
quadratic nonlinear media can support four-hump vortex solitons
with unit topological charge that are stable provided that their
propagation constant is above a certain critical value. Below this
critical value we have identified two types of instabilities: (i)
an exponential-type of instability leading to the final decay and
spread out of the solitons across the lattice and (ii) an
oscillatory-type instability leading to the transformation of the
lattice vortex soliton into a fundamental soliton without internal
vorticity. We investigated the generation of the multicolor
lattice vortex soliton from Gaussian beams with nested phase
dislocations. The possibility to generate different output lattice
soliton patterns, with and without vorticity, by varying the
topological charges and amplitudes of the input beams in seeded
excitation configurations, has been discussed. The generation of a
2D periodic potential in quadratic nonlinear media is a
challenging issue, even though fabrication of 1D lattices has been
already achieved using techniques which might be extended to 2D
geometries. Also, the results presented here might be relevant to
suitable atomic-molecular Bose-Einstein condensates held in
optical lattices.

This work was partially supported by the Generalitat de Catalunya,
by the Instituci\'{o} Catalana de Recerca i Estudis Avan{\c c}ats
(ICREA), and by the Spanish Government through grant BFM2002-2861.


\newpage

\textbf{Figure Captions}

Fig. 1. (a) Profile and (b) phase of vortex solitons supported by
the harmonic lattice at $b_{1}=1.07$. (c) Profile and (d) phase of
vortex soliton at $b_{1}=2$. Only the FF wave is shown. Lattice
depth $p=4$, phase mismatch $\beta=0$.

Fig. 2. (a) Vortex soliton energy flow versus propagation constant
for different values of phase mismatch at $p=8$. (b) Propagation
constant cutoff versus lattice depth at $\beta=0$. (c) Cutoff
versus phase mismatch at $p=8$. (d) Stability and instability
domains for different lattice depths at $\beta=0$. Circles show
critical value of propagation for stabilization.

Fig. 3. Propagation of vortex solitons with $b_{1}=3.1$ (a), $3.4$
(b), and $5$ (c) in the presence of input noise with variance
$\sigma_{noise}^{2}=0.01$. FF wave profile is shown at different
propagation distances. Lattice depth $p=8$, phase mismatch
$\beta=0$.

Fig. 4. Snap-shot images showing decay of the stable vortex
solitons caused by removal of the lattice. Only SH wave profile is
shown. Images are taken after each 2.5 propagation units. Lattice
depth $p=8$, phase mismatch $\beta=0$.

Fig. 5. Generation of the vortex solitons with only FF input. (a)
Field and (b) phase distributions of the input FF beam with
topological charge $m_{1}=1$. (c) FF beam and (d) SH beam at
$\xi=15$. Lattice depth $p=8$, phase mismatch $\beta=0$.

Fig. 6. Soliton algebra. The output soliton distribution depends
on the topological charges $m_{1}$ of FF wave and $m_{2}$ of SH
wave, respectively. In all cases, $m_{1}=1$. In (a)-(d), the
amplitude of FF wave $A=20$ and the amplitude of SH wave $B=2$. In
(e) and (f) $A=20$ and $B=0.5$. Plots (a)-(f) correspond the
topological charges $m_{2}=1,3,4,6,7,8$ respectively and show the
output SH field distribution at $\xi=100$. Lattice depth $p=8$,
phase mismatch $\beta=0$.

\end{document}